\documentclass[%
 amsmath,amssymb,
 aps, b
 prl,
 twocolumn,
 superscriptaddress,
 floatfix
]{revtex4-2}

\usepackage{graphicx}% Include figure files
\usepackage{bm}% bold math
\usepackage{color}
\usepackage{soul}
\usepackage[normalem]{ulem}
\usepackage{hyperref}

\begin{document}

\preprint{APS/123-QED}

\title{Crystallization and topology-induced dynamical heterogeneities in soft granular clusters}% Force line breaks with \\

\author{Micha{\l} Bogdan}
\email[]{mbogdan@ichf.edu.pl}
\affiliation{Institute of Physical Chemistry, Polish Academy of Sciences, Kasprzaka 44/52,
01-224 Warsaw, Poland}

\author{Jesus Pineda}
\email[]{jesus.pineda@physics.gu.se}
\affiliation{Department of Physics, University of Gothenburg, Origovägen 6 b
41296 Göteborg, Sweden}

\author{Mihir Durve}
\affiliation{Center for Life Nano- \& Neuro-Science, Fondazione Istituto Italiano di Tecnologia (IIT), viale Regina Elena 295, 00161 Rome, Italy}

\author{Leon Jurkiewicz}
\affiliation{Institute of Physical Chemistry, Polish Academy of Sciences, Kasprzaka 44/52,
01-224 Warsaw, Poland}

\author{Sauro Succi}
\affiliation{Center for Life Nano- \& Neuro-Science, Fondazione Istituto Italiano di Tecnologia (IIT), viale Regina Elena 295, 00161 Rome, Italy}
\affiliation{Istituto per le Applicazioni del Calcolo del Consiglio Nazionale delle Ricerche, via dei Taurini 19, 00185 Rome, Italy}
\affiliation{Department of Physics, Harvard University, 17 Oxford St., Cambridge, Massachusetts 02138, USA}

\author{Giovanni Volpe}
\email[]{giovanni.volpe@physics.gu.se}
\affiliation{Department of Physics, University of Gothenburg, Origovägen 6 b
41296 Göteborg, Sweden}

\author{Jan Guzowski}
\email[]{jguzowski@ichf.edu.pl}
\affiliation{Institute of Physical Chemistry, Polish Academy of Sciences, Kasprzaka 44/52,
01-224 Warsaw, Poland}

\date{\today}% It is always \today, today,

\begin{abstract}

 Soft-granular media, such as dense emulsions, foams or tissues, exhibit either fluid- or solid-like properties depending on the applied external stresses. Whereas bulk rheology of such materials has been thoroughly  investigated, the internal structural mechanics of finite soft-granular structures with free interfaces is still poorly understood. Here, we report the spontaneous  `crystallization' and `melting' inside a model soft granular cluster---a densely packed aggregate of $N\sim 30-40$ droplets engulfed by a fluid film---subject to a varying external flow. We develop new machine learning tools to track the internal rearrangements in the quasi-2D cluster as it transits a sequence of constrictions. As the cluster relaxes from a state of strong mechanical deformations, we find differences in the dynamics of the grains within the interior of the cluster and those at its rim, with the latter experiencing larger deformations and less frequent rearrangements, effectively acting as an elastic membrane around a fluid-like core. We conclude that the observed structural-dynamical heterogeneity  results from an interplay of the  topological constrains, due to the presence of a closed interface, and the internal solid-fluid transitions. We discuss universality of such behavior in various types of finite soft granular structures, including biological tissues.

\end{abstract}

\maketitle

%\tableofcontents

Soft granular materials, such as dense foams, emulsions and biological tissues, consist of densely-packed deformable grains, separated by a thin layer of a lubricating fluid \cite{Guevorkian2010, Douezan2011, Manning2010, Cohen-Addad2013, Nezamabadi2017, Kabla2012, Pawlizak2015}. Complex behaviors, like memory effects, avalanches, plasticity and viscoelasticity, emerge from multi-scale interactions in these systems   \cite{Jiang1999, Marmottant2013, Kumar2020, Goyon2008, Goyon2010, Lulli2018, Montessori2021a}.   
Despite advances in soft glass theories of bulk materials  \cite{Sollich1998}, the structural mechanics and micro-rheology of finite-sized soft granular clusters remain poorly understood. Examples of such clusters include cell aggregates like embryos  \cite{Foty1994, Schotz2013}, circulating tumour cell clusters \cite{Au2016}, cell spheroids used in biomimetic  experiments \cite{Tlili2022} and droplet/bubble clusters or jets used as templates in fabrication of porous materials \cite{Highley2019, Constantini2018} .

Freely-floating small soft-granular clusters, with droplets playing the role of grains, have been previously assembled using microfluidics \cite{Guzowski2022, Li2017, Guzowski2015, Lai2022, Huang2011}. However, while very small close-packed clusters (typically $N<10$ droplets) formed by the inner droplets in a larger double emulsion drop have been thoroughly studied in terms of their equilibrium stable and metastable morphologies \cite{Kim2011, Guzowski2015, Adams2018}, little is known about the complex internal dynamics of mesoscale clusters, including their micro-rheology and internal solid-fluid transitions in response to external stresses or upon relaxation. 

Recent microfluidic experiments involving aggregates of living cells or synthetic biomimetic prototissues ($N\sim 10^2$ cells) have revealed intriguing relaxation dynamics ~\cite{Tlili2022, Layachi2022}. However, the challenges associated with cell polydispersity and heterogeneous interactions limited the control over the system at the microscale. Here, we propose a convenient model system: a double emulsion drop encapsulating multiple close-packed and highly monodisperse inner droplets, all confined in a shallow microchannel. The system allows to control generation and manipulation of the droplet-clusters as well as to precisely track the internal droplet rearrangements under external microflows. 

%Despite the substantial body of work highlighting the physical properties of soft granular systems of very large or very small sizes, the understanding of the behaviour of such materials at the mesoscale, under strong confinement \cite{Bogdan2022, Gai2016, Gai2016a, Uchic2004, Tlili2022, Guzowski2022} or surrounded by free interfaces \cite{Bogdan2022, Guzowski2022, Montessori2021a}, is very limited. 

It is known that monodisperse 2D systems feature crystallization, understood here as ordering into a 2D Bravais lattice, usually a hexagonal one. However, so far, crystallization has been experimentally observed only in very large systems, and in systems confined by rigid walls \cite{Holtze2008, Gai2016, Gai2016a, Raven2009, Garstecki2006, Panaitescu2012, Rietz2018}. While some limited numerical work on the structural dynamics of freely floating droplet clusters has been previously reported \cite{Montessori2021a, Tiribocchi2021}, the actual order-disorder transitions in such systems have not been investigated. 

\begin{figure}
    \centering
    \includegraphics[width = 0.48\textwidth]{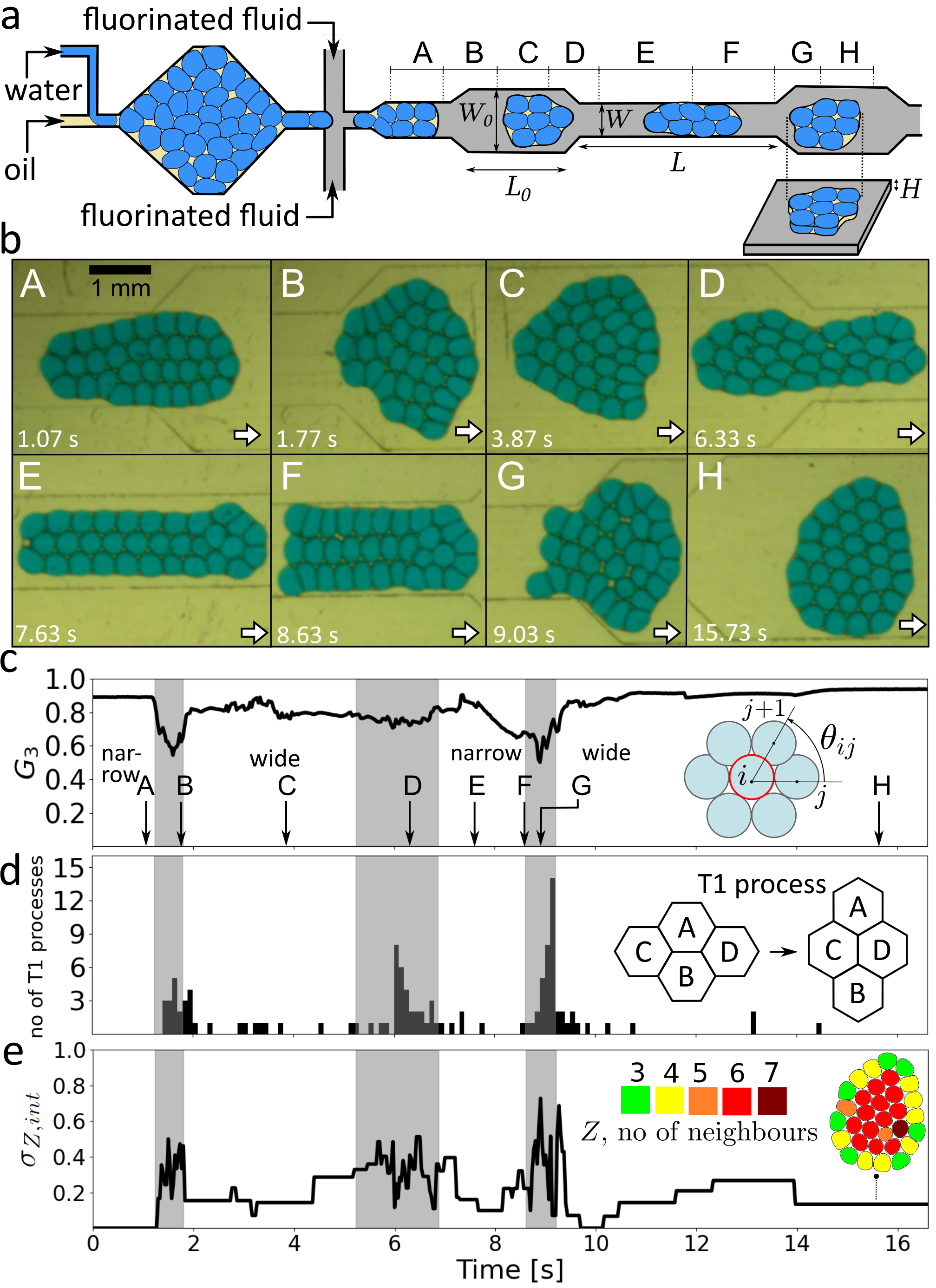}
    \caption{a) Scheme of the microfluidic system for the production of clusters of a soft granular medium. $L = 10 $ mm, $L_0 = 5$ mm, $W = 1.4$ mm, $W_0 = 2.8$ mm, $H = 0.25$ mm. b) Snapshots of the cluster at different times corresponding to different sections of the channel, A-H (see also supporting movie SM1). %A- at rest in the first narrow channel, B- entering the first wide chamber, C- flow within the chamber, D- entering the second narrow channel, E and F- flow through the narrow channel, G- entry into the second wide chamber, H- after a period of relaxation at rest in the wider chamber. 
    c) Order parameter $G_3$ (see main text) of the cluster as a function of time. Gray areas corresponds to the sections  of changing width, white areas - of constant width. d) The number of T1 processes in 0.1 s intervals. e)  The variance $\sigma_{Z,int}$ of the number of nearest neighbors for droplets in the interior of the cluster. The inset displays the color-coded numbers of nearest neighbors for each droplet in the cluster at late times.}
    \label{fig1:system_overview}
\end{figure}

Here, we use a microfluidic system (see Fig. \ref{fig1:system_overview}) which enables sequential generation and manipulation of granular clusters using external flows under quasi-2D confinement. We precisely control the flow of the external carrier fluid to push the clusters via a sequence of constrictions and track the internal dynamics upon flow-induced deformation and subsequent relaxation. 

Our methods are partially based on our previous work, wherein we investigated the stochastic jetting-dripping transition of a close-packed emulsion subject to flow-focusing by an external immiscible phase \cite{Bogdan2022}.
%We have recently developed a method to produce a model soft granular material (a densely-packed double emulsion) and demonstrated that subjecting it to an external flow in a so-called flow-focusing junction leads to novel dynamical modes with highly stochastic phenomena, whose emergence we have causally linked to the granularity of the material 
Here, using a modified experimental setup, we focus strictly on the internal dynamics of the clusters, typically consisting of  $N\sim$ 20-40 droplets. In particular, we observe spontaneously emerging structural-dynamical heterogeneity between two droplet subpopulations: those at the rim of the cluster and those in the interior. To reveal such spontaneous structural segregation we develop a new machine-learning-based method of droplet segmentation and tracking based on the DeepTrack-2 framework \cite{midtvedt2021quantitative, DeepTrack2} (see Supplemental Material for details). We analyse multiple recorded videos frame-by-frame to find that the droplets at the rim get systematically more strongly deformed and rearrange less frequently in response to external flow-induced stresses. Accordingly, we find that the clusters spontaneously develop an internal mesostructure consisting of an elastic granular `membrane' around a more fluid-like granular `core'.

Finally, we also find that sufficiently strong deformations of the cluster, as it transits the constrictions under flow, may eventually force the droplets from the core to enter the rim. This results in effective mixing between the cluster's two subpopulations, which could have applications, e.g., in microfluidics-assisted micro-tissue engineering.

Our double-emulsion system consists of densely packed blue-dyed aqueous droplets encapsulated by a thin film of oil with surfactant (see Supplemental Material for details). The droplets are created at a T-junction and fed to a chamber (see Fig. 1a and SM0). From there, the desired volume of the monodisperse emulsion (dictated by the intended size of the cluster) is pushed out into a narrow channel, where, at a cross-junction, it is cut by a third immiscible phase, a fluorinated liquid. Since the droplets are spontaneously engulfed by the oil phase due to preferential wetting \cite{Guzowski2013}, we observe the formation of a double-emulsion cluster containing multiple aqueous `cores' (droplets) inside a thin oil `shell' carried by the flow of the external fluorinated fluid. The cluster is subsequently pushed through a sequence of widenings and narrowings (see Fig. 1a). We observe the patterns of droplet rearrangements within the cluster upon its compression and relaxation. The velocity of the cluster is regulated by the applied rate of flow of the external fluorinated phase.
We choose a representative cluster ($N=32$, movie SM1) to illustrate a collection of phenomena observed widely throughout the experiments (see Supplemental Material).

We observe crystallization of droplets into a near-hexagonal lattice in the wide fragments of the channel when the clusters move at a constant velocity. The evolution of the cluster in Fig. 1b under an external flow rate of 8 ml/h illustrates this phenomenon. To quantify the level of droplet ordering we analyze the cluster-averaged orientational order parameter $G_3$ \cite{Montessori2021} defined for each droplet $i$ as $G_{3,i} = \frac{1}{N_i}\sum^{N_i}_{j=1}|\textrm{cos}(3 \theta_{ij})|$, where $N_i$ is the number of neighbours of the droplet $i$ and $\theta_{ij}$ is the bond angle centered at droplet $i$ with arms pointing to the centers of two droplets neighbouring droplet $i$ and each other ($j$ labels such pairs, see the inset in Fig. 1c). $G_3$ is known to be the measure of hexagonal order \cite{Montessori2021}, in particular, $G_3 = 1$ for a perfect hexagonal lattice  (Fig. 1c). We also observe that the local arrangements of droplets during transfer through the wider sections remain mostly unchanged, with only individual T1 processes  \cite{Petit2015} (i.e. swapping of neighbours between droplets; see the inset in Fig. 1d for illustration). Note that apart from a classical swapping of neighbours between 4 droplets (as shown in the inset in Fig. 1d), we also consider a T1 process a rearrangement in which a droplet enters or leaves the rim, therefore affecting the order of droplets in the rim. Later in the text we refer to the former type of rearrangement as a `T1 process in the interior' and to the latter as a `T1 process in the rim'.  
As the cluster enters a widening or narrowing (see Figs. 1b- 1c), we observe a rheological solid-fluid transition. This is signified by the increased frequency of T1 processes occurring in parallel with the change of the shape of the cluster. We refer to this process as `melting'. The internal hexagonal structure is destroyed and $G_3$ correspondingly decreases.

The cycles of crystallization and melting can be also seen in the variance $\sigma_{Z,int}$ of the number $Z$ of neighbours of droplets within the interior of the cluster (Fig. 1e), where, as expected for a monodisperse system, median($Z$) = 6. 
%For completeness, we also show $\sigma_{Z,rim}$  for the rim population (median($Z$) = 4). 
We find that $\sigma_{Z,int}$ is relatively low in channel sections of a constant width, or in the absence of flow, and increases when the cluster transits a constriction. %For topological reasons, a subset of rim droplets, in regions of high interface curvature, have 3 neighbors instead of 4, which leads to an inherently larger variance. 
In the final relaxed state, we find 1 positive and 7 negative defects in the 'rim' (defined as the droplets directly touching the interface of the emulsion with the external fluid) and 1 positive and 1 negative topological defect in the 'interior' (the remaining droplets). This configuration closely resembles the energy-minimizing cluster ($N=32$) predicted numerically by \citet{Cox2003}. We conclude that the observed cluster relaxes to a final state which is very close to the actual global energy minimum.

%\hl{[the problem with the previous paragraph is that it does not show any difference between core and shell in terms of fluidity, does it?; or maybe we can find something?]}

Overall, we find that the extensional or compressional flows, experienced by a cluster upon entering or exiting a constriction, increase the level of fluidity in the system. This can be also analyzed based on the patterns of droplet motion in the center-of-mass reference frame (see Supplementary Material for a detailed discussion). 

We note that a flow of 8 mL/h (as in SM1) implies a capillary number Ca $\equiv \eta U/\gamma \approx 0.37 \times 10^{-2}$ and Ca $\approx 0.74 \times 10^{-2}$ within the wide and narrow channel, respectively, where $\eta$ is the viscosity of the oil, $\gamma$ is the interfacial tension between the water and oil phases and $U$ is the velocity of the fluids, averaged over the channel cross-section. We note that this is close to the values of Ca at which a crossover between fluid-like and solid-like patterns of motion have been reported for a monodisperse quasi-2D dense emulsion \cite{Gai2019}.

\begin{figure}
    \centering
    \includegraphics[width = 0.48\textwidth]{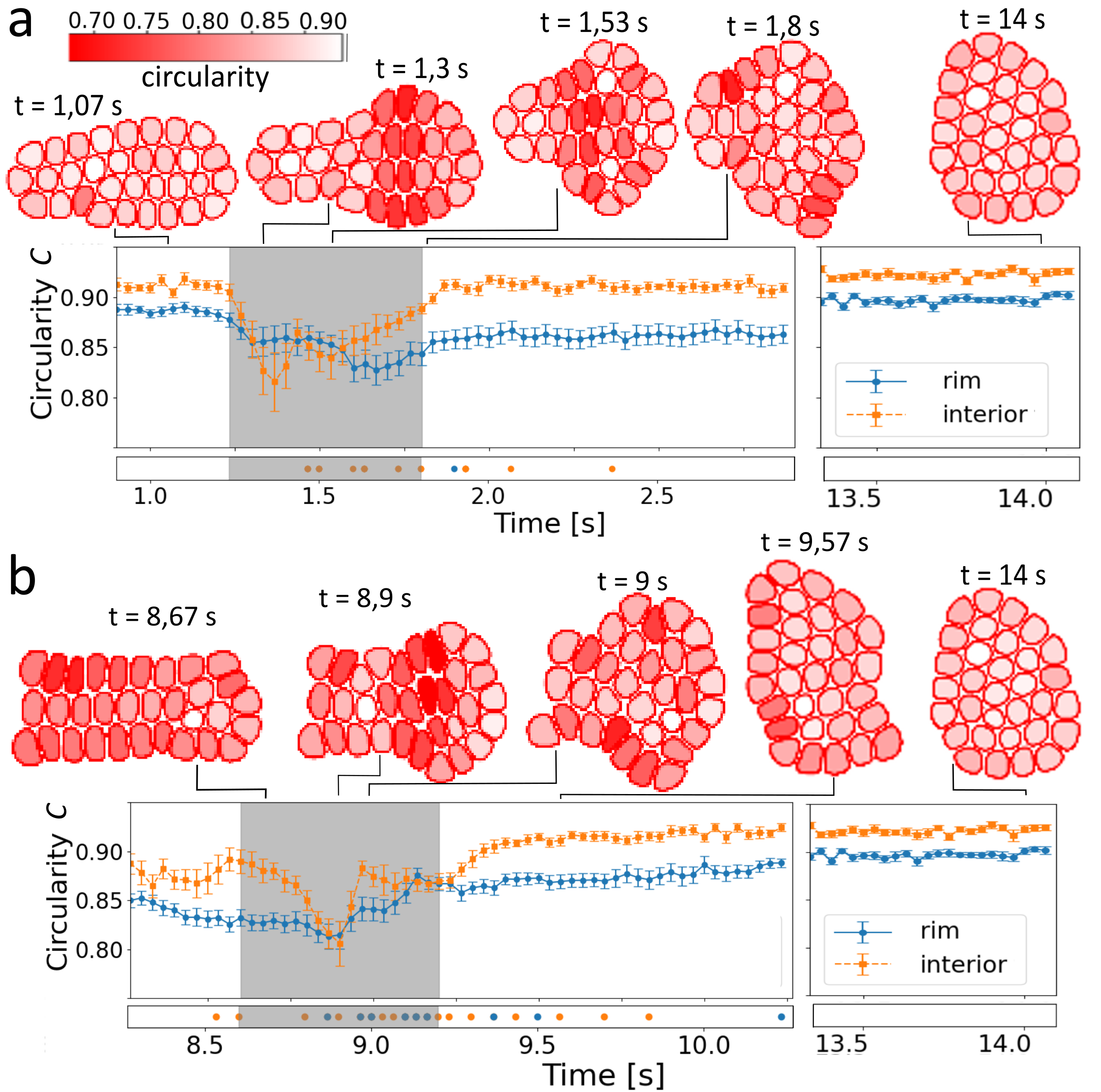}
    \caption{(a) Circularity of the droplets $C \equiv 4 \pi A/P^2$, where $P$ is the perimeter and $A$ the area of a droplet (note $4 \pi A/P^2=1$ for a perfect circle), in the rim and interior subpopulations as a function of time for the first relaxation event (sections A-C of the channel, see Fig. 1a).  The color-coded snapshots display the spatial variations in the droplet circularity (red-lowest, white-highest) within the cluster at different time-points. The separate plot and snapshot on the right indicate the close-to-equilibrium configuration at late times. The occurrence of T1 processes in the rim (blue) and interior (orange) of the cluster (see the main text for a precise definition) is indicated as a point process in time beneath the main plot. (b) Data analogous to (a) for the second relaxation event (sections F-H). Error bars indicate the standard errors of the mean.}
    \label{fig:my_label2}
\end{figure}

One of our most intriguing findings is that the process of melting, understood as the intensification of T1 processes, occurs mostly in the interior of the cluster, while the rim of the cluster responds more as an elastic solid, absorbing stresses via droplet deformations (see Figs. 2a-2b and SI text). In particular, we observe that, upon exiting a constriction, the droplets in both interior and rim undergo fast, externally-enforced shape deformations, and develop a temporary compression front which gradually dissolves as the droplets in the interior start to rearrange undergoing numerous T1 processes and quickly regaining their average circularity $C \equiv 4 \pi A/P^2$, where $P$ is the perimeter and $A$ the area of a droplet (note $4 \pi A/P^2=1$ for a perfect circle). In contrast, the droplets in the rim feature fewer T1 processes and recover their circular shape much more slowly. Quantitatively, during the relaxation stage (see $t\in [1.36, 4.13]$ s and $[8.85, 15.73]$), whose onset is defined as the point when $C$ of the interior droplet population reaches a minimum, we observe around 5 times less rearrangements at the rim than in the interior of the cluster. Based on 6 relaxation events and 149 T1 processes in movies SM1, SM2, SM3, SM4 and SM5, we obtain an average 0.31 T1 processes per droplet in the rim and an average 1.52 T1 processes per droplet in the interior per relaxation event. We suspect that this `dynamic segregation' effect results from stronger confinement of the droplets at the rim and their resultant inability to release mechanical frustrations. 

\begin{figure}
    \centering
    \includegraphics[width = 0.48\textwidth]{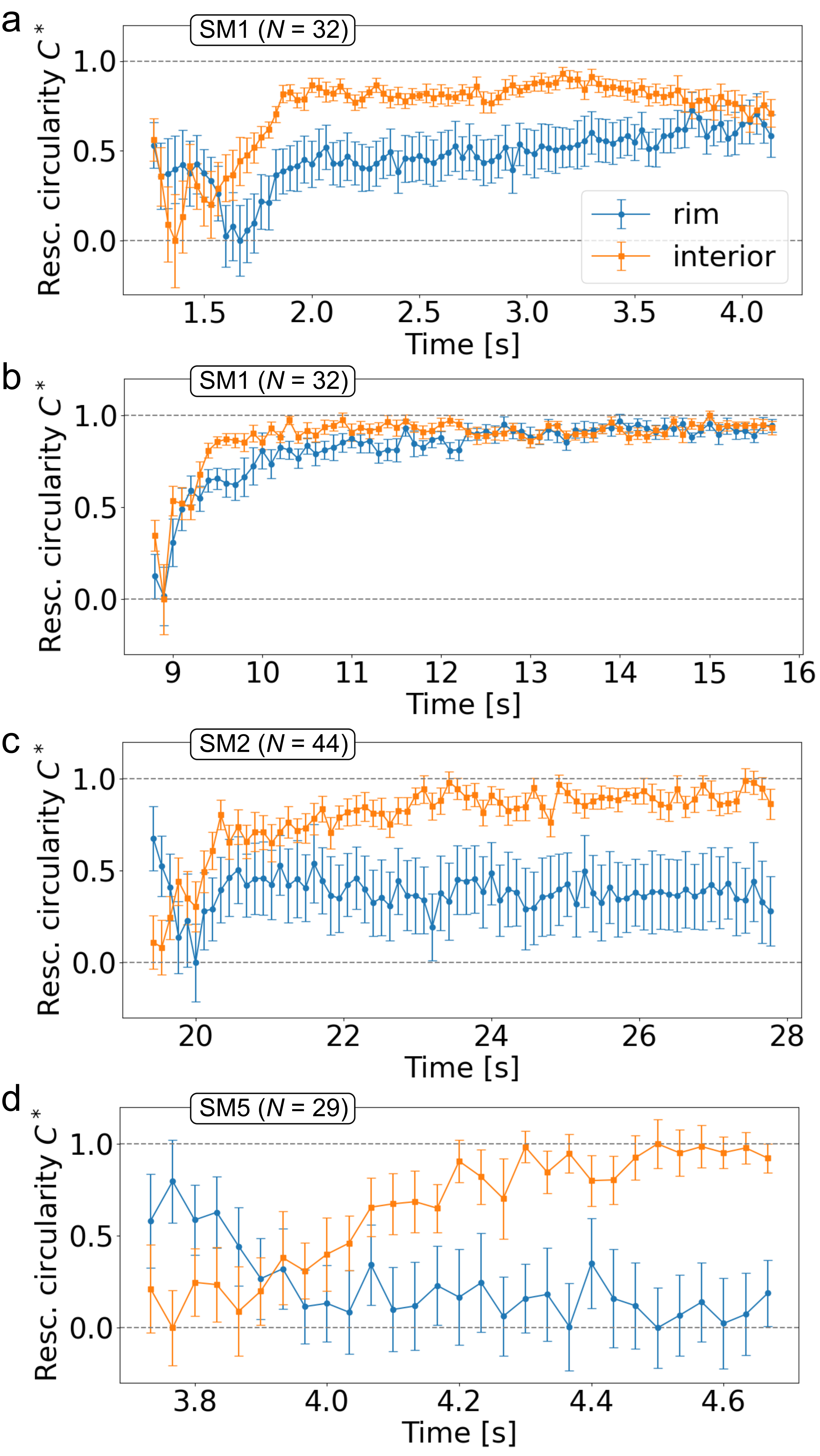}
    \caption{Rescaled circularity of the droplets defined as $C^*(t) \equiv (C(t) - C_{min})/(C_{max} - C_{min})$ in the rim and interior populations during the relaxation phase for 3 different clusters (from top to bottom respectively clusters in supporting movies SM1, SM1, SM2, SM5). Panels a and b correspond to different compression events for the same cluster. Error bars indicate the standard errors of the mean.}
    \label{fig:my_label3}
\end{figure}

We note that the expected \emph{equilibrium} circularity values for the interior and the rim populations are actually different due to the slightly different droplet shapes in the corresponding inner and outer regions of the cluster. This can be seen in Fig.\ \ref{fig:my_label2}, where the initial and late-time values of $C$---which may be considered very close to the global equilibrium---differ by around $3\%$ for the two populations. In fact, the droplets at the rim remain slightly more deformed than in the interior. The difference may seem minor, however, it is comparable with the overall variations in the circularities during the relaxation process, the latter not exceeding around 10$\%$. Therefore, in order to properly visualize the relaxation dynamics for the two droplet populations, in Fig.\ \ref{fig:my_label3}a-b we re-plot the respective circularities shifted with respect to the minimum value and normalized to the overall amplitude, i.e., we plot $C^*(t)=(C(t)-C_{min})/(C_{max}-C_{min})$. Note that $C_{min}$ is taken as the minimum of $C(t)$ over the chosen relaxation event while $C_{max}$ is taken as the maximum of $C(t)$ over the whole experiment (not only the relaxation phase). With such a choice, $C_{max}$ serves as the best approximation of the actual equilibrium. The rescaled plots in Fig.\ \ref{fig:my_label3} directly demonstrate the effect of the \emph{dynamic} heterogeneity of the cluster, i.e., with subtracted effects of the \emph{static} (equilibrium) heterogeneity. The effect of the dynamic segregation is clearly reproducible for sufficiently large clusters ($N\sim 30-40$; see Fig.\ \ref{fig:my_label3}c-d). For smaller clusters ($N\sim 20$) we observe more stochastic behavior and less (or none) segregation effects (see Figs S3c-S3d and S5). 

In the case of the larger clusters, in 3 out of 4 considered relaxation events (Fig.\ \ref{fig:my_label3}a,c-d), we also find that the interior population experiences the maximal reduction of circularity before the rim population. This characteristic 'phase-shift' in the early stages of the relaxation period signifies that the rapid fluidization of the interior accompanies initial solid-like deformation of the rim. At later stages, in all cases (Fig.\ \ref{fig:my_label3}a-d), the rim population recovers its circularity much slower than the interior. Note that in the event presented in Fig.\ \ref{fig:my_label3}d, due to the short observation window, the rim population remains close to the maximally distorted state up to the end of the plotted period, while the interior quickly relaxes and reaches the maximum circularity. Note also that for clarity in panels (a) and (b) we have plotted $C^*(t)$ for every third and every second frame respectively.

In summary, for apparently mainly topological reasons, under moderate perturbations, the relaxing clusters develop a fluidized `core' surrounded by an elastic `membrane'. We propose that the effect occurs because the shape of the cluster is changed by the external flow at a timescale $\tau_{flow}$ which is shorter than or comparable to the timescale $\tau_{T1}$ needed for a T1 process to occur. To check this hypothesis, we measure $\tau_{T1}$ during periods when the cluster is flowing in the wider chamber without extensional strains ($t\in [2.53, 4.90]$ s in SM1). Based on averaging over 6 T1 processes we obtain $\tau_{T1} = 0.56$ s. We also estimate $\tau_{flow} \approx l/(U - U_0)$, where $U=Q/(HW)$,  $U_0=Q/(HW_0)$  are the average velocities of the flow in the narrower and wider sections of the channel, respectively, and $l = 0.7$ mm is the length over which the width of the channel is changing. We find $\tau_{flow} = 0.22$ s, such that  $\tau_{flow}\lesssim \tau_{T1}$, in agreement with the expectations.  We also note that, for the segregation effect to be observable, the timescale of the experiment $\tau_{exp}$ must also be much longer than $\tau_{T1}$. Accordingly, the general condition can be formulated as   $\tau_{flow}\lesssim \tau_{T1}\ll \tau_{exp}$. It is obeyed in our case, since $\tau_{exp}\simeq 13$ s and $\tau_{exp}\in [4, 15]$ s in the other experimental runs.

In general, the arrest of the droplets at the rim could be caused not only by the topology, i.e., the closed interface, but also by direct interaction of the droplets with the interface, e.g., due to the partial wetting of the droplets by the external, fluorinated phase. In our case, however, the three-phase (water-oil-external) contact angle appears to be too small to be measured directly, from which we conclude that the adhesion is likely negligible. Accordingly, we argue that the arrest of the droplets at the interface is mainly of topological origin whereas the interface effectively reduces the number of degrees of freedom and prevents, or significantly slows down, the rearrangements. Nevertheless, in general, the prevalence and relative importance of either the `topological arrest' or direct adhesion to the interface would depend on the actual experimental conditions, e.g., the precise chemical compositions of the external phase and the middle `oil' phase. 

%Further studies examining the consequences of changing these parameters would be in order.

\begin{figure}
    \centering
    \includegraphics[width = 0.48\textwidth]{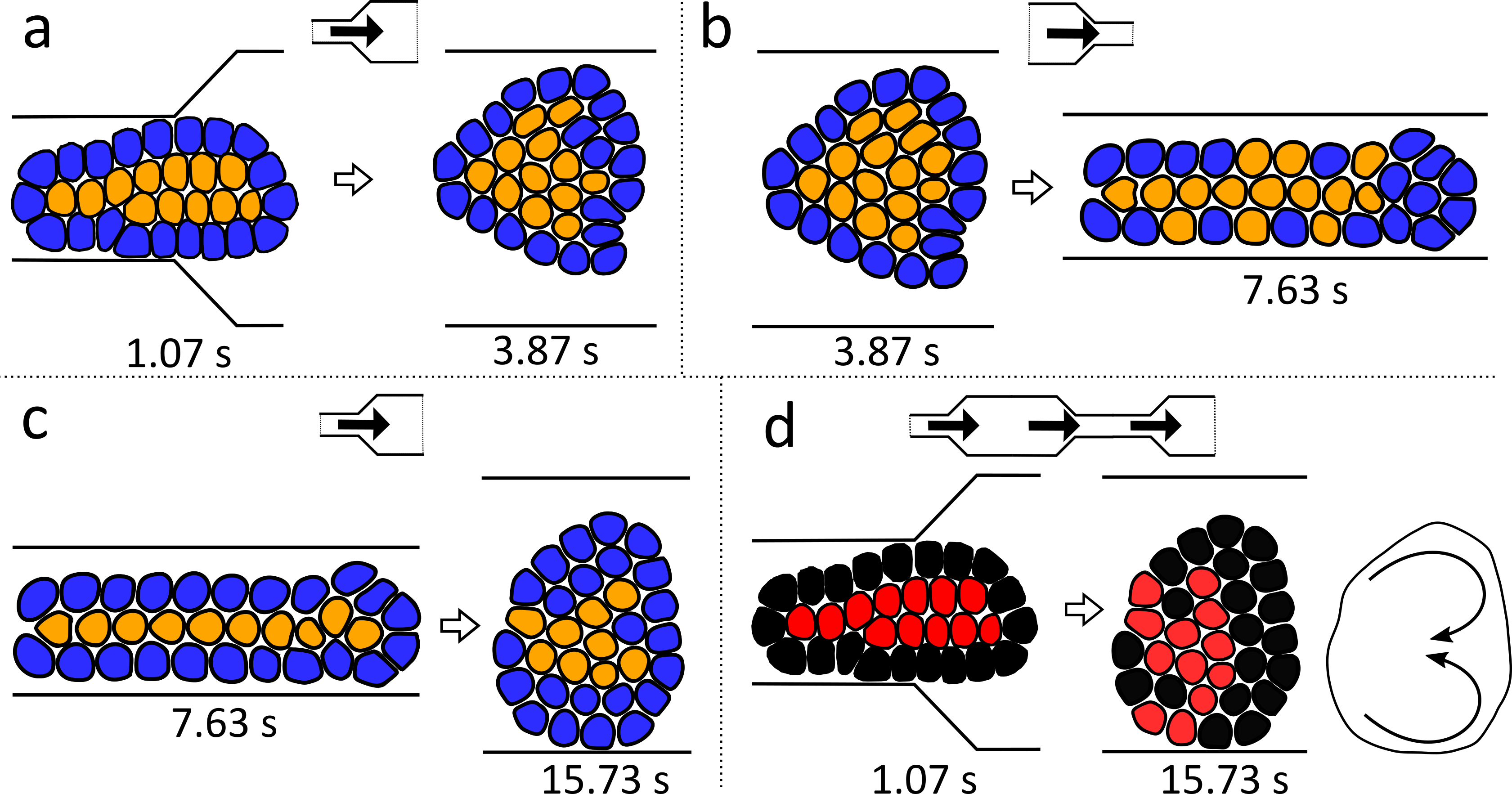}
    \caption{Mixing of droplets forming the initial rim (blue) and interior subpopulations (yellow), separately for each of the  three translocations: (a) first widening, (b) narrowing, (c) second widening. (d) Cumulative process for the entire sequence, with initial rim  and interior droplets colored black and red, respectively. The scheme on the right indicates the direction of effective re-circulation.}
    \label{fig:my_label4}
\end{figure}

We also find that, in spite of the enhanced arrest of the droplets at the rim during the relaxation events, sufficiently strong deformations of the cluster upon transiting the subsequent constrictions can be used to impose gradual mixing of the rim and interior subpopulations (Fig. \ref{fig:my_label4}). In fact, we find that as the cluster exits the first constriction, it rounds up and reduces its perimeter (Fig. \ref{fig:my_label4}a, $t=$ 3.87 s), so that eventually some droplets from the frontal, downstream section of the rim enter the core. Subsequently, as the cluster enters the second constriction (Fig. \ref{fig:my_label4}b), the frontal part is again 'rolled up' forcing even more droplets from the rim into the interior. Later in the constriction, as the cluster unfolds into a 3-row 'plug'  ($t=$ 7.63 s) and increases its perimeter, the droplets from the interior enter the rim, whereas upon exiting the constriction (Fig. \ref{fig:my_label4}c) we observe an opposite process. Overall, however, there is no reversibility in the droplet rearrangements (Fig. \ref{fig:my_label4}d) and we observe gradual mixing within the cluster. The process apparently involves shifting of the droplets from the front to the back of the cluster, thus resembling internal re-circulation in a simple viscous plug \cite{Zhu2016}, however, with the opposite direction of the re-circulation.

%In a previous study \cite{Bogdan2022}, we have shown that a soft granular material flowing under strong confinement features new dynamical modes and that the chaotic behaviour of individual grains often determines the properties of the entire system. 

In summary, we report the microrheology and internal structural transitions within soft granular clusters modeled by a double-emulsion drop with multiple close-packed inner droplets. We find that in the mid-size clusters ($N\sim 30-40$) the subpopulations of the droplets within the interior and at the rim of the cluster develop different mechanical responses to the external stresses, with the rim acting effectively as an elastic membrane around a more fluid-like core, an effect which we attribute to the topology of the system itself (a closed interface).
%, since otherwise the droplets are indistinguishable (do not differ in content, size, etc). 

Due to the topological nature of this effect, it could be present in a wide spectrum of soft granular materials with free interfaces in the situations when they approach the fluid-solid transition, such as e.g. cell aggregates, spheroids or organoids subject to external stresses. The degree to which the effect would occur may, in general, depend on the details of grain-grain and grain-interface interactions, as well as on the relation between the timescale associated with the external stresses (flow in our case) and the timescale of T1 rearrangements. 

Regarding the interactions, one may expect the phenomenon to hold in 'homogeneous' cases, with all grains being equivalent. For example, in our case, the elastic repulsive interactions can be considered homogeneous throughout the cluster. Regarding the timescales, we provide a short comparison with the work of Tlili et al. \cite{Tlili2022} who studied translocation of a cluster of cells through a long, narrow microchannel. The authors observed individual cell elongations, but virtually no T1 processes. The reason for that was that  $\tau_{exp}$ was much smaller than $\tau_{T1}$. In fact, the authors measured $\tau_{T1}$ in a separate long-term experiment in which only several neighboring cells were tracked, but not the whole cluster. Therefore, the dynamic effects associated with cell rearrangements, such as interior-rim separation, were not observed in the translocation experiments.

Nevertheless, in systems with homogeneous interactions and with the proper separation of timescales, we expect the reported dynamics to remain universal. In particular, our results should also translate to 3-dimensional systems and, as such, should be taken into account when considering the effective surface tension and viscosity of cell aggregates and tissues, the parameters of significant interest, e.g., in developmental biology and tissue engineering \cite{Steinberg2007, Schotz2008, Stirbat2013, Foty1994}. Differential deformation of cells within spheroids and embryos 
%in comparison to their more circular peers within the interior 
has been observed previously \cite{Manning2010, Xiong2014, Sart2017}. The relative role of the dynamical-topological effects in proliferating systems remains to be investigated.

At present, a comprehensive theory predicting the onset of fluidization, crystallization and effective dynamical heterogeneity in soft granular clusters, going beyond the qualitative arguments outlined above, is missing. A possible approach would encompass applying the kinetic theory of flow in soft glassy materials \cite{Bocquet2009}, which implies a non-local constitutive law for flow coupled to the rate of plastic events. Such theory remains yet to be adapted to systems with a closed, free boundary. The type of the boundary conditions could be empirically established based on experimental results such as ours, and finally the system could be solved numerically. 
%The time-dependent boundary conditions could be empirically established based on our experimental results. Such a system could in principle be solved numerically. To improve the model, a discretization of the system into separate domains, corresponding to droplets, could be envisaged. 
We leave such theoretical investigations as a future work of ours or others.

Our study opens up avenues for further experimental research. For example, the behavior of soft granular clusters at junctions or in extensional flows, resulting in stretching and breakup, remain poorly understood, and could be studied with a modified version of our system.

Finally, in terms of applications, we find that narrow orifices imposing significant elongation of a cluster can potentially serve as efficient mixers for microscale granular media. The observation could be exploited to control cell rearrangements, e.g., within cell  spheroids, with potential applications in  tissue- and organoid-engineering.
%e.g. to achieve the effect of partially displacing invasive leader cancer cells from the rim to the bulk of a tumour spheroid on metastatic efficiency \cite{Cheung2013, Hallou2017}. 
Such studies could also help understand mechanical principles of tissue flow which remain of great relevance not only in tissue engineering \cite{Kosztin2012}, but could also shed light on the basic mechanisms of wound healing \cite{Petitjean2010} and/or cancer progression including metastasis \cite{Friedl2009a}.

\subsection*{Acknowledgments}

The authors acknowledge funding from the European Union’s Horizon 2020 research and innovation programme under the Marie Skłodowska-Curie grant agreement No. 847413 and from the European Research Council (ERC-PoC grant No. 101081171 (DropTrack) and grant No. 101001267 (MAPEI) under the Horizon Europe research and innovation funding programme). M.B. acknowledges the PMW programme of the Minister of Science and Higher Education in the years 2020-2024 no. 5005/H2020-MSCA-COFUND/2019/2 and support from the National Science Center (Narodowe Centrum Nauki) within the Miniatura grant no. 2022/06/X/ST3/01504.
J.G. acknowledges the support from the National Science Center within Sonata Bis program under Grant No. 2019/34/E/ST8/00411. M.B., L.J. and J.G. thank Patryk Adamczuk for technical assistance. M.D. and S.S. thank Adriano Tiribocchi and Marco Lauricella for fruitful discussions. We gratefully acknowledge the HPC infrastructure and the Support Team at Fondazione Istituto Italiano di Tecnologia.

\subsection*{Author contributions}

M.B., S.S., G.V., and J.G. conceived and designed the study. M. B. and L. J. performed the experiments. M. B., J. P., M. D., L. J. and J. G. analyzed the data. M. B., J. P., M. D. and L. J. contributed materials and/or analysis tools, M.B., J. P., M.D., S.S., G. V. and J.G. wrote the manuscript.

\bibliography{main}% Produces the bibliography via BibTeX.

\end{document}

\preprint{APS/123-QED}

\title{Supplemental Material for the article \textit{Crystallization and topology-induced dynamical heterogeneities in soft granular
clusters}}% Force 

\author{Micha{\l} Bogdan}
\email[]{mbogdan@ichf.edu.pl}
\affiliation{Institute of Physical Chemistry, Polish Academy of Sciences, Kasprzaka 44/52,
01-224 Warsaw, Poland}

\author{Jesus Pineda}
\email[]{jesus.pineda@physics.gu.se}
\affiliation{Department of Physics, University of Gothenburg, Origovägen 6 b
41296 Göteborg, Sweden}

\author{Mihir Durve}
\affiliation{Center for Life Nano- \& Neuro-Science, Fondazione Istituto Italiano di Tecnologia (IIT), viale Regina Elena 295, 00161 Rome, Italy}

\author{Leon Jurkiewicz}
\affiliation{Institute of Physical Chemistry, Polish Academy of Sciences, Kasprzaka 44/52,
01-224 Warsaw, Poland}

\author{Sauro Succi}
\affiliation{Center for Life Nano- \& Neuro-Science, Fondazione Istituto Italiano di Tecnologia (IIT), viale Regina Elena 295, 00161 Rome, Italy}
\affiliation{Istituto per le Applicazioni del Calcolo del Consiglio Nazionale delle Ricerche, via dei Taurini 19, 00185 Rome, Italy}
\affiliation{Department of Physics, Harvard University, 17 Oxford St., Cambridge, Massachusetts 02138, USA}

\author{Giovanni Volpe}
\email[]{giovanni.volpe@physics.gu.se}
\affiliation{Department of Physics, University of Gothenburg, Origovägen 6 b
41296 Göteborg, Sweden}

\author{Jan Guzowski}
\email[]{jguzowski@ichf.edu.pl}
\affiliation{Institute of Physical Chemistry, Polish Academy of Sciences, Kasprzaka 44/52,
01-224 Warsaw, Poland}

\date{\today}

\maketitle

%\tableofcontents

\section*{\label{sec:materials_and_methods} Experimental methods}

The microfluidic chip was fabricated in polycarbonate via milling. Additional PDMS copies of the chip (closed with a glass slide) were produced by using the polycarbonate chip as the master (see \cite{Postek2017} for details of the procedure). Both polycarbonate and PDMS versions of the chip were used in the experiments. A hydrophobic modification of the chips was achieved by the injection of 3M's Novec 1720 Electronic Grade Coating into the channels, which was later allowed to air dry in temperatures of up to 120 °C in the polycarbonate version of the chip and up to 75 °C in the PDMS version of the chip. In the case of observed wetting of the walls of the channels by the oil, the hydrophobic modification was repeated.

The droplets within the clusters consist of water dyed with 0.1\% w/w Erioglaucine. The oil used as the lubricating phase is a 1:1 mixture of 5 cSt  Xiameter PMX-200 silicone fluid and hexadecane, with added surfactant (2\% w/w SPAN80) warranting the stability of the droplets (note the relative proportions of the 5 cSt  Xiameter PMX-200 silicone fluid and hexadecane was around 7:3 in our previous work \cite{Bogdan2022}). The fluorinated fluid used to cut the water-in-oil emulsion into clusters is FC40 with 0.5 \% w/w PFPE-PEG-PFPE fluorosurfactant. The interfacial tension $\gamma$ of the water-oil interface is 3.4 mN/m, as measured by the pendant drop method. Based on literature values, the viscosity of the oil $\eta$ is estimated as 4.0 mPa s.  The flow of all three phases is controlled using a high-precision syringe pump (Cetoni Nemesys 290 N).
The motion of the clusters is recorded using a uEye camera. The frame rates in frames per second (fps) for respective movies are as follows: 36 fps for SM0, 30 fps for SM1, SM2, SM3, SM4 and 35 fps for SM5.

\section{Droplet segmentation and tracking}

\begin{figure*}[hbt!] 
    \centering
    \includegraphics[width=1\textwidth]{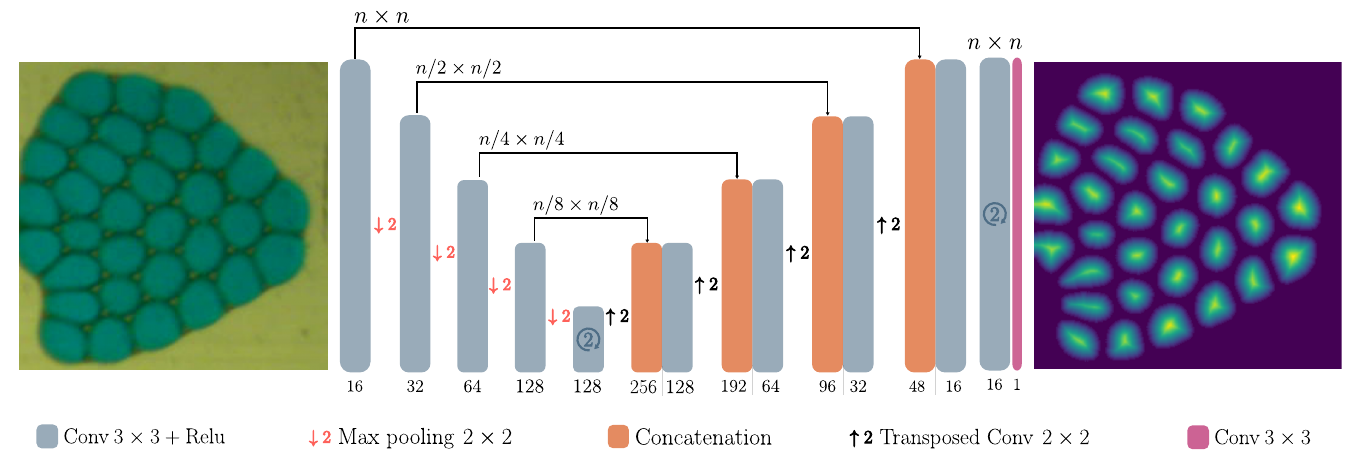}
    \caption{\textbf{Overview of the deep-learning model employed for droplet segmentation}. The neural network architecture is based on the UNet framework. The encoder comprises convolutional blocks followed by max-pooling layers with a stride of 2 for downsampling. Conversely, the decoder employs transposed convolutions for upsampling and incorporates concatenation layers and convolutional blocks. The model leverages skip connections to transfer data between corresponding layers in the downsampling and upsampling paths. The depth of the neural network has been designed to facilitate an optimal receptive field for analyzing larger droplets. Importantly, the neural network is designed to output a distance transform image, wherein every pixel encodes the Euclidean distance from its position to the nearest background pixel. Following training, a binary classification can be obtained by applying a threshold to the network output.}
    \label{fig:neural_network}
\end{figure*}

\begin{figure*}[hbt!] 
    \centering
    \includegraphics[width=1\textwidth]{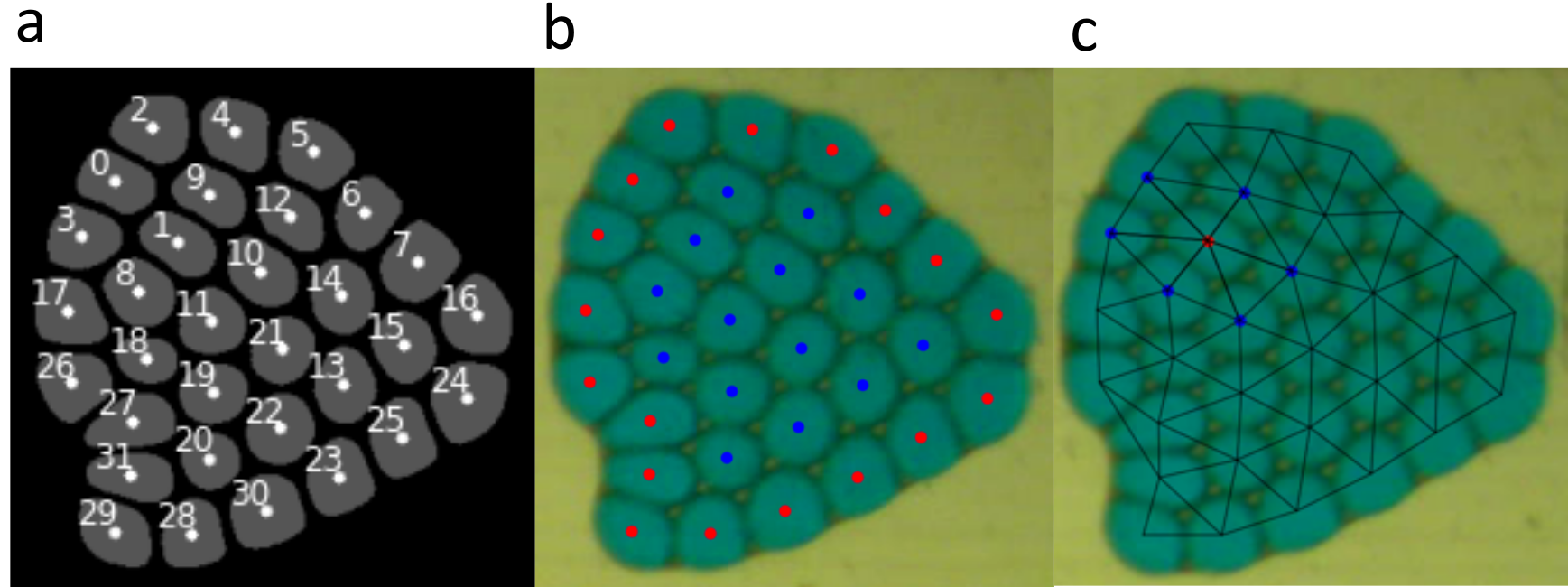}
    \caption{\textbf{Effects of higher-level data analysis on segmentation data}. (a) Droplet identification for tracking (b) Identification of droplets as belonging to the "rim" or "interior" (c) Identification of droplet neighbours.}
\label{fig:data_analysis}
\end{figure*}

A deep learning-based approach is employed to accurately segment the droplets. The droplets are segmented and tracked using the deep learning framework DeepTrack-2~\cite{midtvedt2021quantitative}. The network architecture employed is similar to the UNet~\cite{ronneberger2015}, comprising down-sampling and up-sampling paths, with skip-connections in between (Fig.~\ref{fig:neural_network}). Conventionally, UNet-like models are trained to directly output masks representing different objects in the image. However, to achieve the desired segmentation, multiple post-processing stages with heuristics are often necessary. Inadequate post-processing can result in segmentation errors, especially when dealing with touching objects such as droplets. To overcome these limitations and enhance the neural network performance, our UNet considers a different approach by predicting the Euclidean distance from each pixel to the nearest background pixel (known as distance transform) instead of performing a pixel-wise classification of the input image. This approach has proven to offer several advantages, such as allowing accurate segmentation of touching droplets, facilitating the detection of large droplets, and reducing learning complexity, even with a relatively small dataset~\cite{grauer2021active, sierra2023corneal}.

The training process follows a three-stage approach. Firstly, an initial dataset is created using the Segment Anything package (SAM)~\cite{kirillov2023segment}, which serves as an unbiased annotation alternative. SAM is applied to a single representative experimental video to generate annotations encompassing various droplet shapes and configurations. Subsequently, the initial annotations undergo refinement by a human expert. This refinement involves meticulously removing partially annotated or over-segmented droplets, ensuring higher annotation quality and accuracy. This stage aims to improve the dataset by eliminating inconsistencies and errors. 
Finally, the refined segmentations are converted into distance transform images, which are, in turn, utilized to train the neural network (see Fig.~\ref{fig:neural_network}).
Optionally, the second and third stages can be iterated several times until optimal segmentation results are achieved. This iterative approach allows for further improvement in the dataset quality and the overall performance of the neural network. Compared to manual annotations, which are limited by human-level accuracy and lack generalizability, the proposed training process provides a consistent and robust strategy, particularly in situations where a labeled dataset is unavailable. Once trained, the neural network generalizes over previously unseen videos without losing precision.

During training, we use the backpropagation algorithm to iteratively optimize the neural network's trainable parameters~\cite{rumelhart1986learning}. The training goal is to minimize the mean squared error (MSE) between the predicted and ground-truth distance transform images. Moreover, we utilized the Adam optimizer~\cite{kingma2014adam} with a learning rate of 0.0001 and trained the network for 100 epochs. Each epoch consisted of 1024 randomly selected $128 \times 128$ image patches, divided into batches of 32. The training samples were augmented using rotational, mirroring augmentations, and motion blurring.

Once the droplets are segmented, the subsequent step in our tracking method involves the construction of track segments by linking the individual droplet detections. Specifically, we employ the Linear Assignment Problem (LAP)\cite{jaqaman2008robust}, a mathematical optimization algorithm that facilitates the association of elements from two sets, in this case, linking detected droplets across consecutive frames. To implement LAP, a cost matrix is constructed using the Euclidean distance between centroids of potential associations. The Euclidean distance serves as a measure of dissimilarity, allowing LAP to optimize the track segment assignment by minimizing the total cost. Given that our droplets exhibit well-defined separation, we found the droplet centroids to be a dependable representation for our tracking purposes. Notably, only spatiotemporally close detections are considered for association, ensuring that the tracking algorithm focuses on relevant and proximate droplet pairs.
Furthermore, to account for the influence of an external flow on droplet motion, we restrict connections to detections that align with the flow direction. LAP can be directly applied in steady flow conditions. The relatively slow movement of droplets allows for efficient centroid-based tracking. Yet, an additional pre-processing step becomes essential for accelerating flow dynamics. In such scenarios, the rapid motion of the entire droplet cluster surpasses the internal rearrangement of individual droplets. Taking advantage of this discrepancy, we spatially align the time subsequent clusters before applying LAP. Doing so ensures LAP operates on a coherent representation of the cluster's motion, mitigating the occurrence of mistaken associations in track segments. This tracking algorithm is used to obtain data on flow patterns within the cluster (see Fig. \ref{fig:flow_patterns}).

\section{Additional data analysis}

Higher-level droplet properties have been investigated using data analysis techniques performed on the droplet areas provided by the segmentation algorithm (see Fig. \ref{fig:data_analysis}). Droplets have been classified as belonging to the "rim" or "interior" based on the proportion of pixels within their bounding boxes belonging to the "background" area as obtained by the segmentation algorithm (see Fig. \ref{fig:data_analysis}b). A subsequent visual verification determined this algorithm provided results with essentially no incorrect attributions in the analyzed movies serving as bases for figures presented in the main text (Fig. 2) or SI (Figs. \ref{fig:circularity1} and \ref{fig:circularity2}). The determination of droplet neighbours was carried out based on a Delaunay triangulation of the set of center points of the droplets (see Fig. \ref{fig:data_analysis}c). The algorithm produced only individual incorrect attributions, which were subsequently corrected manually in the resultant dataset used to produce Fig. 1e of the main text.

Since the measurement of the frequency of T1 processes is very sensitive to even infrequent errors in automatic neighbour attribution, while the rearrangements are relatively easy to spot by eye, the occurrences of T1 processes are marked manually in Fiji \cite{Schindelin2012}, with data exported to .csv files. To compare the frequencies of T1 processes during the relaxation stage in the rim and the interior of the cluster, the following time ranges, corresponding to relaxation events, were chosen: SM1, $t \in [1.36, 4.13]$ s; SM1, $t \in [8.85, 15.73]$ s; SM2, $t \in [19.46, 28.00]$ s; SM3, $t \in [8.30, 10.70]$ s; SM4, $t \in [8.60, 12.43]$ s; SM5, $t \in [3.77, 4.7]$ s. 

%The number of droplets within the rim and interior evolves slightly during each of these events. Since we are testing the hypothesis that droplets within the rim are less likely to rearrange than droplets within the interior, we always take the maximum of

\section*{\label{flow} Rim vs. interior droplet heterogeneity in additional events}

We use the system to produce around 20 droplet clusters, apart from the representative cluster described in the main text. Several additional clusters were chosen to extract data, based on having a sufficient size (at least 20 droplets) and monodispersity of the droplets (with a measured CV of droplet cross-section areas below 20\%). A detailed statistical comparison between large populations of clusters may be carried out in future works. The basic parameters of these clusters are presented in table  \ref{tab:basic_cluster_data}.

\begin{table}
  \centering
  \begin{tabular}{|c|c|c|}
    \hline
    Cluster & No of droplets & Flow rate [ml/h] \\
    \hline
    SM1 & 32 & 8 \\
    \hline
    SM2 & 44 & 4 \\
    \hline
    SM3 & 20 & 32 \\
    \hline
    SM4 & 22 & 8 \\
    \hline
    SM5 & 29 & 8 \\
    \hline
  \end{tabular}
  \caption{Basic parameters of clusters}
  \label{tab:basic_cluster_data}
\end{table}

\begin{figure*}
    \includegraphics[width = \textwidth]{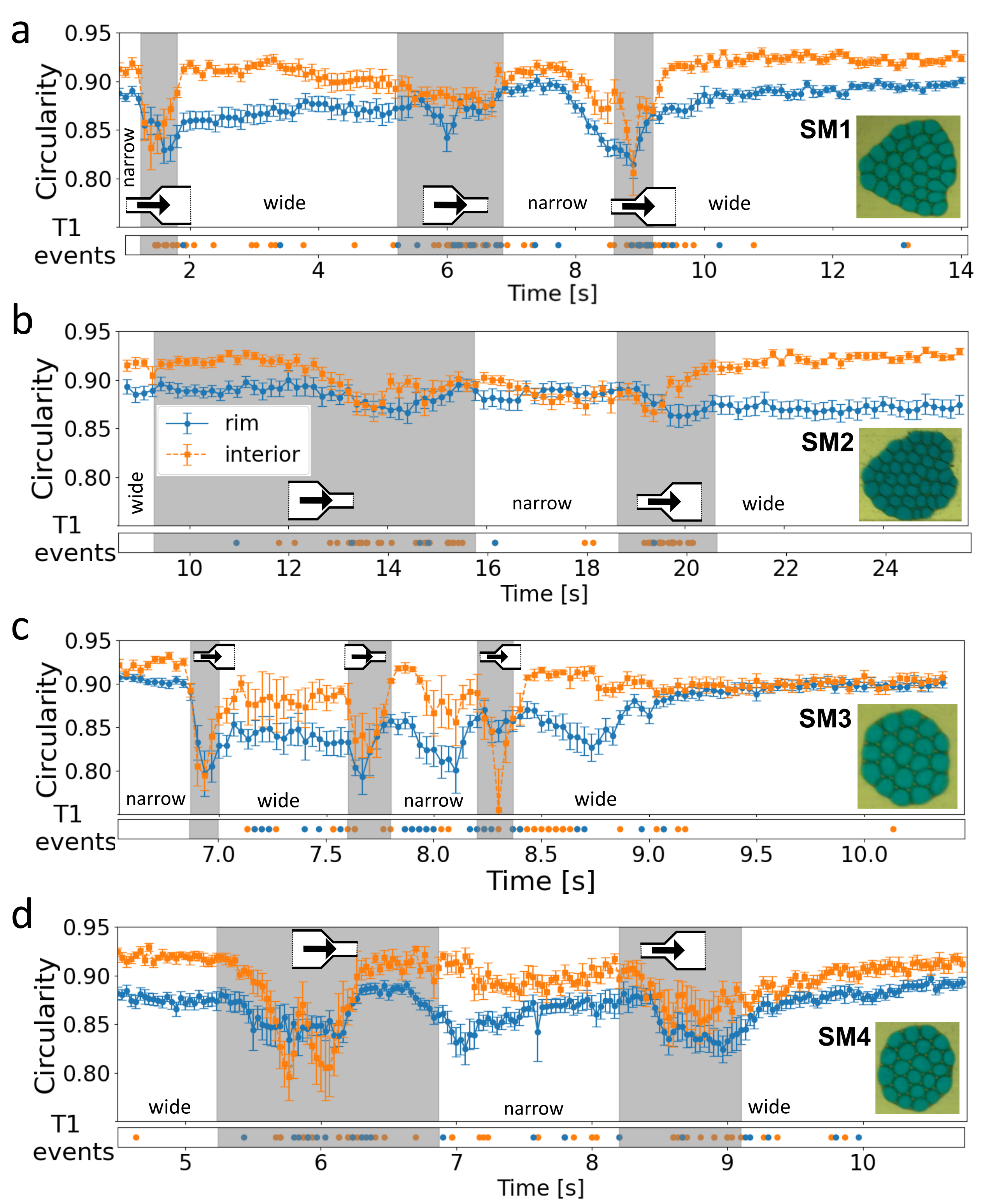}
    \caption{(a, b, c, d) Circularity $4 \pi A/P^2$ (as defined in the main text) of rim (blue) droplets and interior (orange) droplets for 4 clusters visualised in movies SM1, SM2, SM3 and SM4 respectively. Error bars indicate the standard errors of the mean. The lower panel in each subplot marks the occurrence of T1 processes in the rim and the interior of the cluster. The gray area of the plot corresponds to moments in which the cluster is going through fragments of the channel with changing width.}
    \label{fig:circularity1}
\end{figure*}

\begin{figure*} 
    \includegraphics[width = \textwidth]{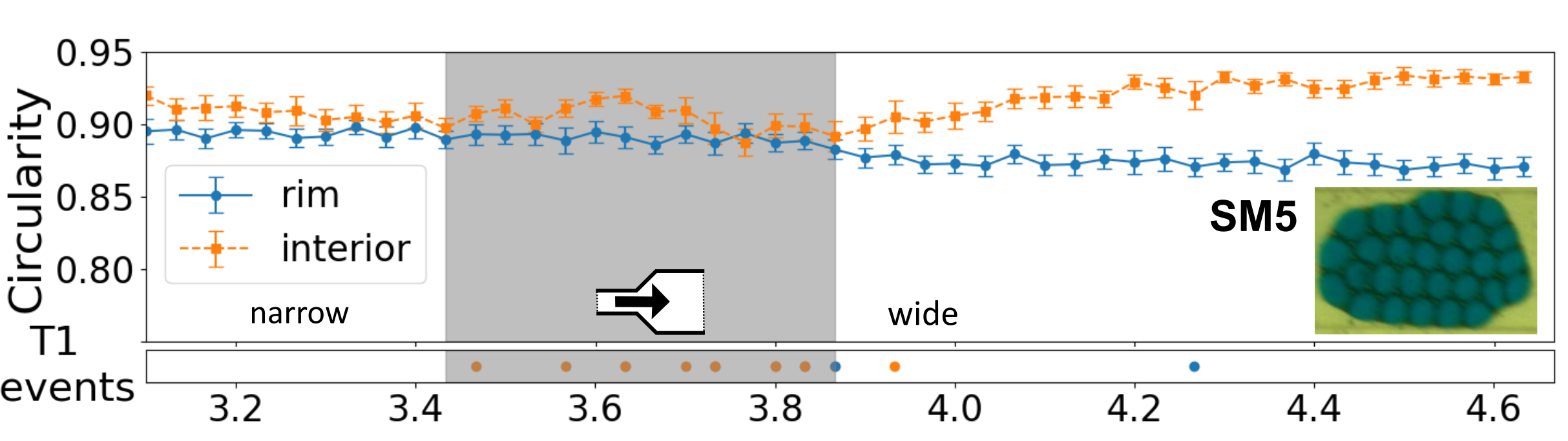}
    \caption{Circularity $4 \pi A/P^2$ (as defined in the main text) of rim (blue) droplets and interior (orange) droplets for the cluster visualised in movie SM5. Error bars indicate the standard errors of the mean. The gray area of the plot corresponds to moments in which the cluster is going through fragments of the channel with changing width.}
    \label{fig:circularity2}
\end{figure*}

Figs. 2a- 2b of the main text demonstrate the effects of the cluster in movie SM1 moving from a narrow channel to a wider chamber on droplet elongations and the frequency of T1 processes. These plots are discussed in the main text. Here, in Figs. \ref{fig:circularity1} and \ref{fig:circularity2}, we present data based on 5 clusters visualised in movies SM1- SM5 (we recommend watching respective supporting movies alongside reading the plots). Figs. \ref{fig:circularity1}a-\ref{fig:circularity1}d of this SM demonstrate changes in circularity and frequencies of T1 processes occurring as each of the respective clusters goes through a cycle of compression as it enters a narrow channel from a wide chamber, and then subsequent relaxation after entering the second wide chamber (in Figs. \ref{fig:circularity1}a and \ref{fig:circularity2}c, an additional previous relaxation event when the cluster enters the first wide chamber was also filmed, with results also presented in the figure). 

Fig. \ref{fig:circularity1}a demonstrates the entire history of changes in the cluster presented in the movie SM1. It is apparent that the difference in circularity between the rim and interior droplets rapidly increases when the cluster exits the constriction and later gradually decreases, ultimately reaching a level corresponding to the state of mechanical equilibrium (apparently being very close to the global energy minimum as we discuss in the main text).

Fig. \ref{fig:circularity1}b paints a similar picture for the cluster visualised in movie SM2, which is both larger than the cluster described in the main text and moves two times slower. Like in the case of the previous cluster, we observe that the relatively small initial difference in circularity between the interior and rim droplets initially decreases inside the constriction but then rapidly increases as the cluster exits the constriction.

Fig. \ref{fig:circularity1}c demonstrates the evolution of circularity of the relatively small cluster (20 droplets) in movie SM3 as it transits two constrictions. Similar to the previous cases, we observe rapid rise in the circularity difference upon the cluster exiting both constrictions. In addition, we observe a similar effect when the cluster enters the constriction. It is worth noting that this cluster is pushed by an external flow twice as large as the cluster in SM1.

Fig. \ref{fig:circularity1}d shows the only cluster in which the trend visible in all other clusters is not quite observed. It is worth noting that, due to the relatively small size of this cluster (22 droplets) the previous interior of the cluster gets almost entirely incorporated into the rim as the cluster translocates the orifice, which may be why the results observed in larger clusters are not replicated here. 

\begin{figure}
    \centering
    \includegraphics[width = 0.48\textwidth]{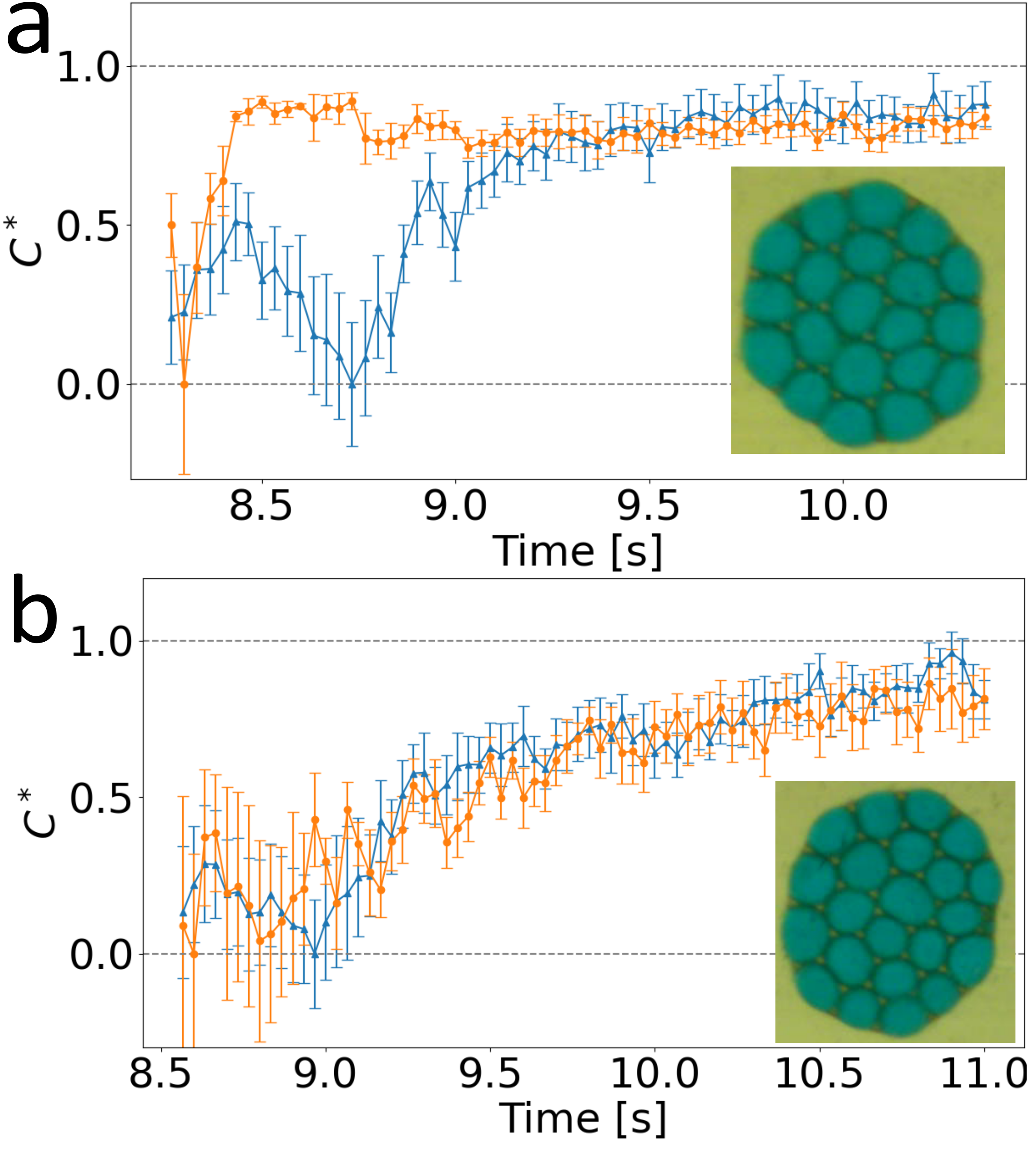}
    \caption{Re-scaled circularity of the droplets defined as $C^*(t) \equiv (C(t) - C_{min})/(C_{max} - C_{min})$ in the rim and interior populations during the relaxation phase after 2 different compression events for 2 relatively small clusters (from top to bottom respectively clusters in supporting movies SM3, SM4). Error bars indicate the standard errors of the mean. Note that for clarity in panels (a) and (b) we have plotted $C^*(t)$ for every third and every fifth frame respectively.}
    \label{fig:my_label_S_rescaled}
\end{figure}

Fig. \ref{fig:circularity2} shows analogous results for an additional cluster, but in this case only a single relaxation event was captured. In consistency with the previous clusters, the initially very moderate difference in circularity between the core and the rim is quite durably increased after the cluster moves from a narrowing to a wide chamber, with many T1 processes occurring in the core, but only one in the rim. It is worth noting that the observation of the relaxation of the cluster was much slower than in other cases. Figs. \ref{fig:my_label_S_rescaled}a and \ref{fig:my_label_S_rescaled}b show the evolution of the rescaled circularity $C^*(t)$, as defined in the main text, for the clusters SM3 and SM4 after they leave the second orifice. Note that the dynamic segregation effect is visible in the case of the cluster SM3, yet not in the case of SM4.

Movies SM1a, SM1b, SM2a, SM3a, SM4a and SM5a show graphically how droplet circularity in specific regions of the respective clusters changes as the clusters leave the orifice and relax (the color scale is in each case identical to the one in Fig. 2 of the main text).  

It must be noted that, as lower panels in each of the subplots demonstrate, the clear majority of T1 processes occurs within the interiors of the clusters, as  opposed to their rims; this is despite the fact that the rim populations tend to be similar in size or larger than the interior populations, depending on the moment in a cluster's evolution.

All in all, while a certain level of stochasticity and individuality can be observed for each cluster and event, the overall picture is that both droplets within the rim and interior of the cluster become elongated due to fast, externally enforced, changes in the shape of the cluster, but the droplets in the interior are able to relax the mechanical frustrations much faster via T1 processes. Droplets in the rim are forced to maintain their frustrated shapes for a longer time. This effectively leads to a dynamic phase separation into a fluid-like core and an elastic solid-like rim of the cluster.

\section*{\label{flow} Flow fields in soft granular clusters}

\begin{figure*}
    \includegraphics[width = 0.99 \textwidth]{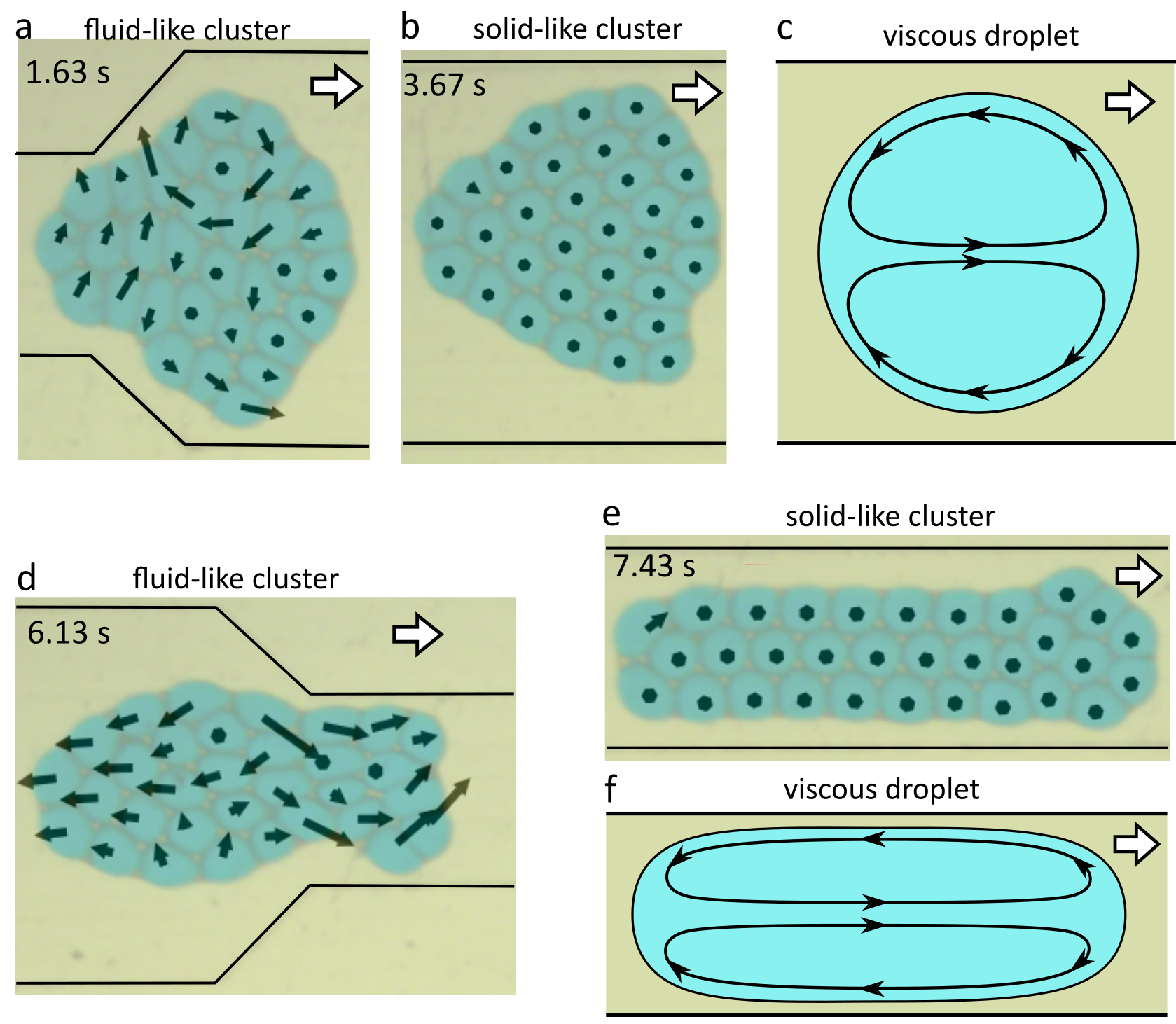}
    \caption{(a,b,d,e) Velocities of individual droplets relative to the cluster's center of mass in movie SM1. Each arrow's length is proportional to the displacement of the droplet within 0.167 s after the snapshot. (c,f) Flow re-circulation patterns in viscous droplets schematically shown for comparison, based on \cite{Zhu2016}.}
    \label{fig:flow_patterns}
\end{figure*}

The level of fluidity/solidity in the system can also be analysed based on the patterns of droplet motion within the cluster. Fig. \ref{fig:flow_patterns} shows the instantaneous velocities of droplets of the cluster relative to its center of mass at several selected snapshots. At the entry to the first widening (Fig. \ref{fig:flow_patterns}a) the cluster features local velocity swirls, with many droplets moving towards the upper or lower wall of the channel, often in a direction differing from that of their neighbours. Once the cluster is flowing within the widening, it largely moves like a rigid, solid body (Fig. \ref{fig:flow_patterns}b), but as it enters the narrowing, it becomes stretched, with the front and rear of the cluster moving in opposite directions (Fig. \ref{fig:flow_patterns}d). In all of the frames, the patterns of motion, however, are reminiscent of plastic deformations rather than of viscous flows, which feature a re-circulation pattern as shown in Figs. \ref{fig:flow_patterns}c and \ref{fig:flow_patterns}f \cite{Zhu2016}.

Inside the narrow channel (Fig \ref{fig:flow_patterns}e), while the cluster largely flows as a rigid body, various modes of motion can emerge and destabilize the shape of the cluster (note the squeezing of the droplets, accompanied by some T1 processes; Fig. \ref{fig:circularity1}a, $t\in [7.63,8.63]$ s). This demonstrates, again, that weak flows (as encountered in the wide channel) enable crystallization and solidification of the cluster, while excessive and/or accelerating flows induce internal rearrangements.

%In particular, a cluster flowing through the narrow channel typically experiences gradual build up and release of internal stresses. This can be observed in SM3 (SM3, $t\in [6,7.5]$ s) where some of the compression is released by a re-circulation pattern in which droplets at the rear of the cluster overtake those in the middle (SM3, $t \in [7.50,8.17]$ s). 
%This is likely related to the fact that within the constriction the external phase flows faster than the cluster \cite{Jakiela2012} and therefore pushes at its rear.

%Note the direction of the re-circulation pattern is opposite to what would happen in a viscous droplet \cite{Jakiela2012}, exemplifying the fact that soft granular clusters do not consistently display conventional behaviour of either the "solid" or "fluid" categories. A yet another non-trivial pattern is observed in the cluster visualized in SM4, where, in the narrow section of the channel, it displays a tank-treading motion resembling a vesicle in a shear flow \cite{Kantsler2005} (SM4, $t \in [2.20,3.33]$ s).

%The unstable internal droplet arrangement is likely related to the acceleration of the flow within a constriction. 


\bibliography{supplemental}